\documentclass[aps,prd,showkeys,twocolumn]{revtex4}
\usepackage{epsfig}
\usepackage{graphicx}
\usepackage{amsmath,amssymb,amsfonts,latexsym}
\usepackage{graphicx}

\usepackage{epsfig,amssymb,amsfonts,verbatim}

\usepackage{amsmath}
\usepackage{latexsym}
\usepackage{amsfonts}
\usepackage{amssymb}
\usepackage{color}

\def\bfl{\begin{flushleft}}
\def\efl{\end{flushleft}}
\def\bfr{\begin{flushright}}
\def\efr{\end{flushright}}
\def\bc{\begin{center}}
\def\ec{\end{center}}

\def\ba{\begin{eqnarray}}
\def\ea{\end{eqnarray}}
\def\baa#1{\begin{array}{#1}}
\def\eaa{\end{array}}
\def\bw{\begin{widetext}}
\def\ew{\end{widetext}}

\def\text#1{\mbox{#1}}

\Large
\begin{document}



\title{Renormalized ionic polarizability functional applied to evaluate the molecule-oxide interactions for gas-sensing
applications}

\author{Andrew Das Arulsamy}
\email{andrew.das.arulsamy@ijs.si}

\affiliation{Jo$\check{z}$ef Stefan Institute, Jamova cesta 39, SI-1000 Ljubljana, Slovenia}

\author{Kristina Eler$\check{\rm s}$i$\check{\rm c}$}
\affiliation{Jo$\check{z}$ef Stefan Institute, Jamova cesta 39, SI-1000 Ljubljana, Slovenia}

\author{Uro$\check{\rm s}$ Cvelbar}
\affiliation{Jo$\check{z}$ef Stefan Institute, Jamova cesta 39, SI-1000 Ljubljana, Slovenia}

\author{Miran Mozeti$\check{\rm c}$}
\affiliation{Jo$\check{z}$ef Stefan Institute, Jamova cesta 39, SI-1000 Ljubljana, Slovenia}


\date{\today}

\begin{abstract}
Metal-oxide sensors are widely used due to their sensitivities to different types of gaseous, and these sensors are suitable for long-term applications, even in the presence of corrosive environments. However, the microscopic mechanisms with quantum effects, which are required to understand the gas-oxide interactions are not well developed despite the oxide sensors potential applications in numerous fields, namely, medicine (breath-sensors), engineering (gas-sensors) and food processing (odor-sensors). Here, we develop a rigorous theoretical strategy based on the ionization energy theory (IET) to unambiguously explain why and how a certain gas molecule intrinsically prefers a particular oxide surface. We make use of the renormalized ionic displacement polarizability functional derived from the IET to show that the gas/surface interaction strength (sensing sensitivity) between an oxide surface and an isolated gas molecule can be predicted from the polarizability of these two systems. Such predictions are extremely important for the development of health monitoring bio-sensors, as well as to select the most suitable oxide to detect a particular gas with optimum sensitivity. 
\end{abstract}

\keywords{Metal-oxide sensors; Ionization energy theory, Ionic polarizability, Gas-oxide interaction}

\maketitle

\section{Introduction}

There are three well known methods available for one to use and to improve the detection sensitivity of gaseous with metal-oxide sensors~\cite{moon}. The first method relies on increasing the active surface area per volume in contact with gaseous environment with nanostructures~\cite{devi,cvel1,cvel2,wan,kong,kim,dut}. The second method involves Pt, Pd and Ag dopants, which are doped into host sensors~\cite{sch,epi}, while the last one is applicable to sensors that have the properties of varistors, i.e., the ability to increase the intrinsic electrical conductivity of these sensors with an applied d.c. voltage~\cite{moon}. However, these experimental techniques are macroscopic in origin, which can be justified with a phenomenological theory developed based on the Clifford empirical model and the Gibbs-Duhem-like equations to understand the relationship between surface processes and the changes in the conductance~\cite{abb}.

It is to be noted here however, there are models available to describe the mechanisms of gas-sensing based on the ionosorption concept (adsorbtion of charged oxygen-atoms and -molecules) and oxygen vacancies on the oxide surfaces (reduction-reoxidation mechanism)~\cite{gurlo}. A thorough review and discussion on these gas-sensing mechanisms, possible reactions on the surfaces, and types of gas sensors are given by Gurlo and Riedel~\cite{gurlo}. In addition, the reversible carrier-type transition that occurs during the gas-sensing operation have also been discussed by Galatsis et al.~\cite{gala} and Gurlo et al.~\cite{gurlo2,gurlo3}. Galatsis et al.~\cite{gala} used the temperature-enhanced electron conduction as the primary source of reversible carrier-type transition in SnO$_2$ and Fe doped SnO$_2$, while Gurlo et al.~\cite{gurlo2,gurlo3} employed the ionosorption mechanism and energy band bending concept to support their measurements in $\alpha$-Fe$_2$O$_3$. We will address all of these issues in our subsequent article~\cite{oxide} because these models are not related to sensing mechanism of NH$_3$ and NO$_2$ molecules~\cite{gurlo}.    

In this work, we exploit the IET developed recently in order to reveal the microscopic mechanisms (with quantum effects) to unequivocally understand the interactions involved in the gas-oxide physisorption. We will also expose strategies to select a suitable metal-oxide sensor specifically to detect a particular gas with optimum sensitivity. In addition, the proposals also contain microscopic explanations for the gas-sensing mechanisms. These gas-oxide interaction based mechanisms provide the basis for further investigations to develop new analyte-biomarkers for exhaled breath. The importance of this information (sensing mechanisms) have been emphasized and reviewed recently by Gouma et al.~\cite{gouma,pras,uros,gouma2} with respect to crystal structures and surface grain boundaries. In contrast, our proposal deals with the surface and gaseous chemical compositions, and the difference of the electric dipole moment as the origin of interaction between a given oxide and a gas molecule, atom or plasma. The proposal here is general that can be applied to other non-metallic gas sensors discussed in Refs.~\cite{kaur1,kaur2,kaur3,soo1,soo2}. 

The paper is organized as follows. In the following section, theoretical details are given on the renormalized ionic displacement polarizability that takes into account both the electronic and ionic polarization. We also discuss how classical systems are different from the quantum systems within the concept of polarizability. This is followed by rigorous analysis of applying the IET to study the interaction of NO$_2$ and NH$_3$ molecules on WO$_3$ and MoO$_3$ surfaces. Finally, we will conclude that there are four types of molecule-oxide interactions, and only two of them are responsible for effective and sensitive gas-sensing.
     
\section{Theoretical details}

Details on the IET formalism are given in Refs.~\cite{and1,and2,and3,and4}, which enabled us to refine the Feynman's atomic hypothesis~\cite{feyn2}, that is, all things are made of atoms, each with unique discreet energy levels and their energy-level differences (the ionization energies) determine how they attract or repel each other, if they are a little distance apart or squeezed close together, respectively~\cite{and4}. This refined statement leads to one of the core axioms of the IET$-$ the energy-level spacing in real solids is proportional to their constituent atomic energy-level spacing. This axiom defines the ionization energy approximation, which will be used throughout in this work. The stated energy-level spacing can be incorporated into the Schr$\ddot{\rm o}$dinger equation in order to arrive at~\cite{and1,and4}

\begin {eqnarray}
\hat{H}\varphi = (E_0 \pm \xi)\varphi. \label{eq:1}
\end {eqnarray}

The potential operator in this Hamilton operator, $\hat{H}$ is actually a renormalized screened Coulomb potential~\cite{and5} and the details of applying Eq.~(\ref{eq:1}) can be found in Refs.~\cite{and6,and7}. Here, $E_0$ and $\xi$ are the total energy at zero temperature and the energy-level spacing, respectively, of a given oxide. We also stress here that the IET is a theory to develop theoretical concepts to explain the physical and physico-chemical mechanisms that occur in any strongly correlated matter, as evidenced in Refs.~\cite{and1,and2,and3,and4,and5,and6,and7,and8,and9,and10}. In the subsequent sub-section, we will introduce the concept of renormalized ionic polarizability. 

\subsection{Renormalized polarizability} 

In order to capture the polarizability of cations and anions, we will start from the renormalized ionic displacement polarizability equation~\cite{and8}, [which is based on Eq.~(\ref{eq:1})]

\begin {eqnarray}
\alpha_{d} = \frac{e^2}{M}\bigg[\frac{\exp[\lambda(E_F^0 - \xi)]}{(\omega_{\rm{ph}}^2 - \omega^2)}\bigg], \label{eq:2}
\end {eqnarray}  
  
where $\omega_{\rm{ph}}$ is the phonon frequency of undeformable ions, $1/M = 1/M^+ + 1/M^-$ in which, $M^+$ and $M^-$ are the positively and negatively charged ions due to their different polarizabilities. Here, $\lambda = (12\pi\epsilon_0/e^2)a_B$, $a_B$ is the Bohr radius of atomic hydrogen, $e$ and $\epsilon_0$ are the electronic charge and the permittivity of space, respectively. Hence, the diffusion that has the polarizability term (due to quantum mechanical effect) can take deformability of a given cation and/or anion into account. 

However, Eq.~(\ref{eq:2}) does not apply for the diffusion of nanoparticles (classical) in the colloidal system~\cite{raj}, where the non-existence of the energy-level spacing and polarizability effect for these nanoparticles have been correctly justified. In the case of IET, the polarizability of one ion, affects its nearest neighbor, and then its next nearest neighbor, and so on depending on the strength of the polarization, which in turn gives rise to the long-range Coulomb interaction. On the other hand, undeformable (charged or uncharged) nanoparticles always satisfy the first-order-short-range interaction that can be treated with classical formalism as shown by Ganapathy et al.~\cite{raj}. 

The reason why deformability (due to polarizability) does not play a significant role in the atomic epitaxial growth (shown to be similar to colloidal system~\cite{raj}) is due to the volume of the individual nanoparticle, which contains tens, if not thousands of atoms. Such nanoparticles, if they are metallic, then they can be considered as classically charged-nanoparticles (because they are easily charged), on the other hand, if they are non-metallic, then we may approximate them as uncharged classical nanoparticles. In the former case, the polarization (charge accumulation on the surface without deformation) does not involve discreet energy-level spacing due to the properties of free-electrons, whereas, for the latter case, polarizing a non-metallic and non-ferroelectric nanoparticle with tens and hundreds of atoms is negligible, unless in the presence of very large electric fields and/or the nanoparticles are of very small in size.

\begin{table}[ht]
\caption{Averaged atomic ionization energies ($\xi$) for individual ions, and the averaging were carried out based on the valence state for each element. These elements are arranged with increasing atomic number $Z$ for clarity. The unit kJ/mol is adopted for numerical convenience. Note here that there are no experimental atomic ionization energy data available for W$^{>2+}$.} 
\begin{tabular}{l c c c } 
\hline\hline 
\multicolumn{1}{l}{Ion}            &    ~~~Atomic number  & ~~~Valence    & ~~~$\xi$   \\  
\multicolumn{1}{l}{}                &   ~~~$Z$             & ~~~state      & ~~~(kJ/mol)\\  
\hline 

H                                   &  1   					    &  1+      & 1312 \\ 
N                                   &  7	   	  			  &  4+      & 4078 \\ 
N                                   &  7	   	  			  &  1+      & 1402 \\ 
O                                   &  8	   	  			  &  4+      & 4368 \\ 
Mo                                  &  42 					    &  6+      & 3540 \\ 
W                                   &  74 					    &  6+      & $-$  \\ 
 
\hline  
\end{tabular}
\label{Table:I} 
\end{table}

The averaged atomic ionization energies for all the elements considered in this work are given in Table~\ref{Table:I}, in which, the averaging follows Ref.~\cite{and11}. Prior to averaging, all the atomic ionization energies were taken from Ref.~\cite{web}. 

\section{Analysis on gas sensors}

In our previous work~\cite{and11}, we have been (very) careful in averaging the ionization energies for cations in the presence of changing anion (oxygen) concentration. That is why we emphasized that the model only takes the substitutional elements into account. However, the model has been shown to be suitable for systems with defects, clusters and impurities~\cite{and3}. This means that, any form of cation-substitutions in oxides (for example, in cuprates or manganites) can lead to oxygen defects that will lead to changes in the valence (oxidation) states of multi-valent cations, which has been taken into account explicitly~\cite{and3}. In this section, we will take another step forward and show how one can average the ionization energies for cations and anions provided we are comparing systems that are isolated molecules (non-interacting gaseous state). This information will help us to evaluate the sensitivity of different sensors in different gas environments.

Prior to our evaluations, we need to first invoke the strong electron-electron interaction concept based on the IET for two-electron atom and ions. It reads, larger ionization energy of a given system implies stronger Coulomb repulsion between the inner (core) electron and the outer (valence) electron, which will give rise to \textit{weakly screened} (strongly interacting) outer electron. As a result, this outer electron interacts strongly (via Coulomb interaction) with the inner electron to produce large energy-level spacing (large atomic ionization energy). The proof is given in Ref.~\cite{and3}, which has given rise to the refinement of Feynman's atomic hypothesis stated earlier. 

The oxides, WO$_3$ and MoO$_3$ are selected for analysis due to their wide applications in sensors, as well as the availability of systematic experimental results. In particular, these oxides have been exposed to NO$_2$ and NH$_3$, and their sensitivities were reported by Prasad and Gouma~\cite{pras}. It is to be noted here that the first studies on the sensing properties of WO$_3$ and MoO$_3$ were carried out by Shaver~\cite{sha} and Mutschall et al.~\cite{mut}, respectively. We propose here that the physisorption of NO$_2$ and NH$_3$ onto WO$_3$ and MoO$_3$ solid surfaces depends on the ionic and electronic polarizabilities of the gas molecules with respect to the solid surfaces. For example, if we have a reactive surface with dangling bonds and defects such that the surface electrons are strongly polarizable (reactive) than the gas molecules, then the sensitivity could also be enhanced with stronger physisorption. Note here however, these polarizability-dependent physisorption mechanisms can only be true for a certain narrow temperature-window ($T_1$ to $T_2$, $T_2 > T_1$), such that the sensor recovery time will not be long. For $T < T_1$, the recovery time will be longer due to insufficient thermal energy to overcome the strength of physisorption. Whereas, higher temperatures ($T > T_2$) could lead to chemisorption and/or chemical reaction. 

\begin{figure}
\begin{center}
\scalebox{0.4}{\includegraphics{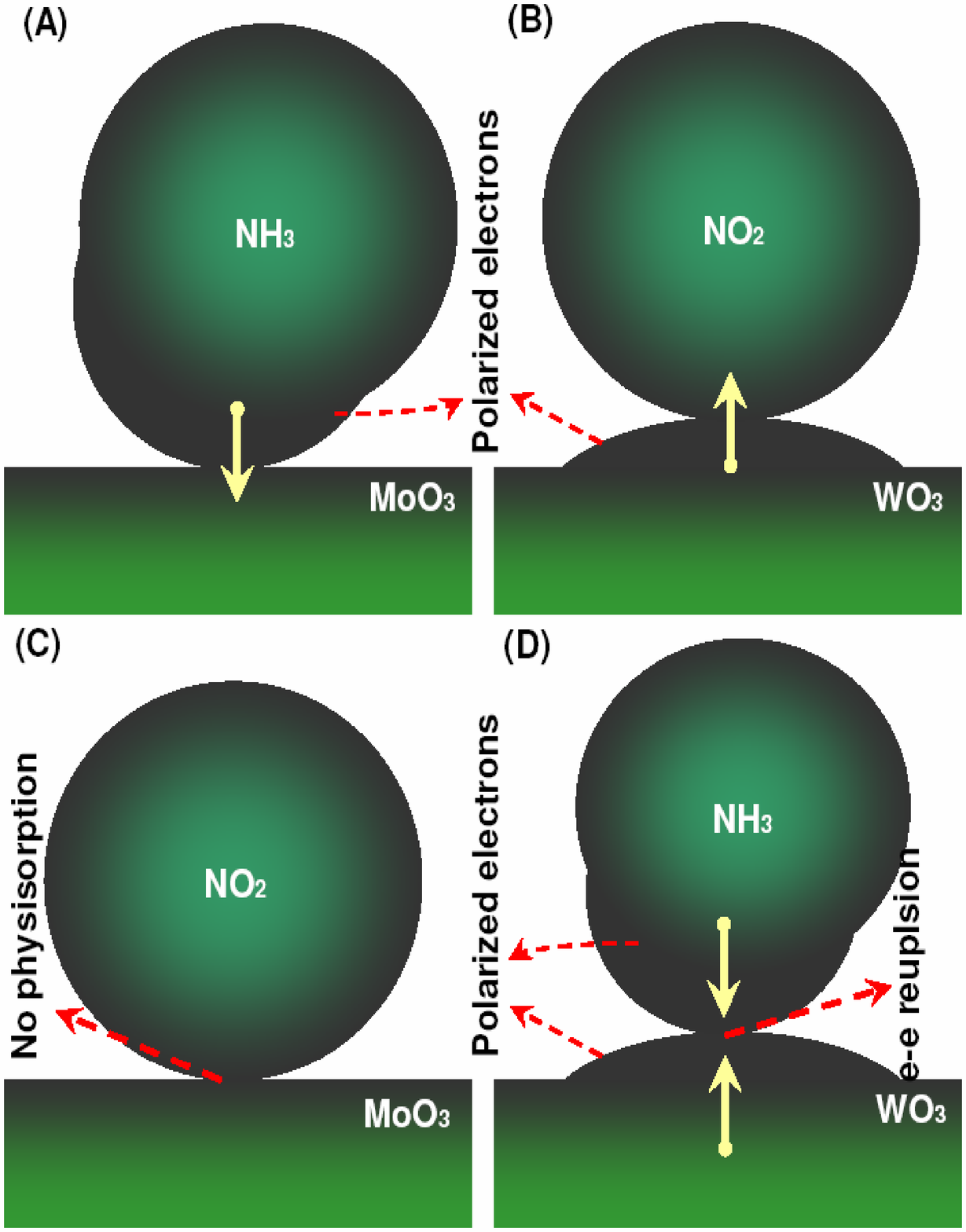}}
\caption{(A) Shows a larger polarizability magnitude for the NH$_3$ molecule in the presence of MoO$_3$ oxide surface, which implies that the electrons transfer from NH$_3$ to MoO$_3$ (follow the solid vertical arrow). (B) WO$_3$ oxide surface has a larger polarizability compared to MoO$_3$ and therefore, the electron-transfer is expected to occur from WO$_3$ to NO$_2$ (also shown with a solid vertical arrow). (C) Unpolarized molecule and oxide surface do not contribute to significant physisorption, and thus, cannot be sensed with any electronic parameter such as resistance. (D) Two strongly polarized entities may repel each other, giving rise to an effective dissociation and negligible detection. The magnitude of the electron-electron ($e$-$e$) repulsion, size of the molecules and the electron polarization are not to scale. See text for details.}
\label{fig:3}
\end{center}
\end{figure}

Apparently, this narrow window depends on the types of solid surfaces and gaseous environments where (a) larger polarizability of a given molecule (NO$_2$ or NH$_3$) will have a stronger physisorption onto the smaller polarizable solid surface, WO$_3$ or MoO$_3$ [see Fig.~\ref{fig:3}(A)]. Conversely, (b) larger polarizable solid surfaces are suitable to detect smaller polarizable gas molecules [see Fig.~\ref{fig:3}(B)]. These two mechanisms originated from the two-electron Coulomb interaction explained earlier~\cite{and3}$-$ electrons from two atoms with large ionization energies (thus, weaker polarization) do not interact to cause detectable change in the conductivity. For example, atomic inert noble gaseous [see Fig.~\ref{fig:3}(C)]. At the other extreme, two highly polarizable atoms will give rise to enhanced dissociation, which is undesirable for effective physisorption because they can be easily evaporated due to polarized electron-electron repulsion~\cite{and5} [see Fig.~\ref{fig:3}(D)]. For example, Si ion on a SiC surface in the presence of electric fields. Here, Si ions (with high polarizability) tend to diffuse and evaporate easily from the SiC substrate compared to C ions (with small polarizability)~\cite{and8}. Of course, these two mechanisms are strictly valid only within that narrow temperature range as pointed out earlier. 


To show that these mechanisms are valid experimentally, we first need to estimate the relative magnitude of the polarizabilities for these gaseous and solid surfaces. In order to do so, we will make use of Eq.~(\ref{eq:2}) that is originated from the renormalized screened Coulomb potential~\cite{and5}. This equation gives us the direct relationship between the constituent atomic ionization energies (types of ions) and the ionic polarizabilities (with renormalized electronic contribution). The strategy here is to find the correct change to the strength of physisorption (due to different polarizabilities), and to determine the sensitivity of different sensors exposed to different gaseous. This information can then be used to predict the temperature range required for optimum sensitivity for different gaseous and sensors. For example, gas molecules with small polarizabilities will need $T_1$ and $T_2$ to be higher than the required range for molecules with large polarizabilities.

\begin{figure}
\begin{center}
\scalebox{0.35}{\includegraphics{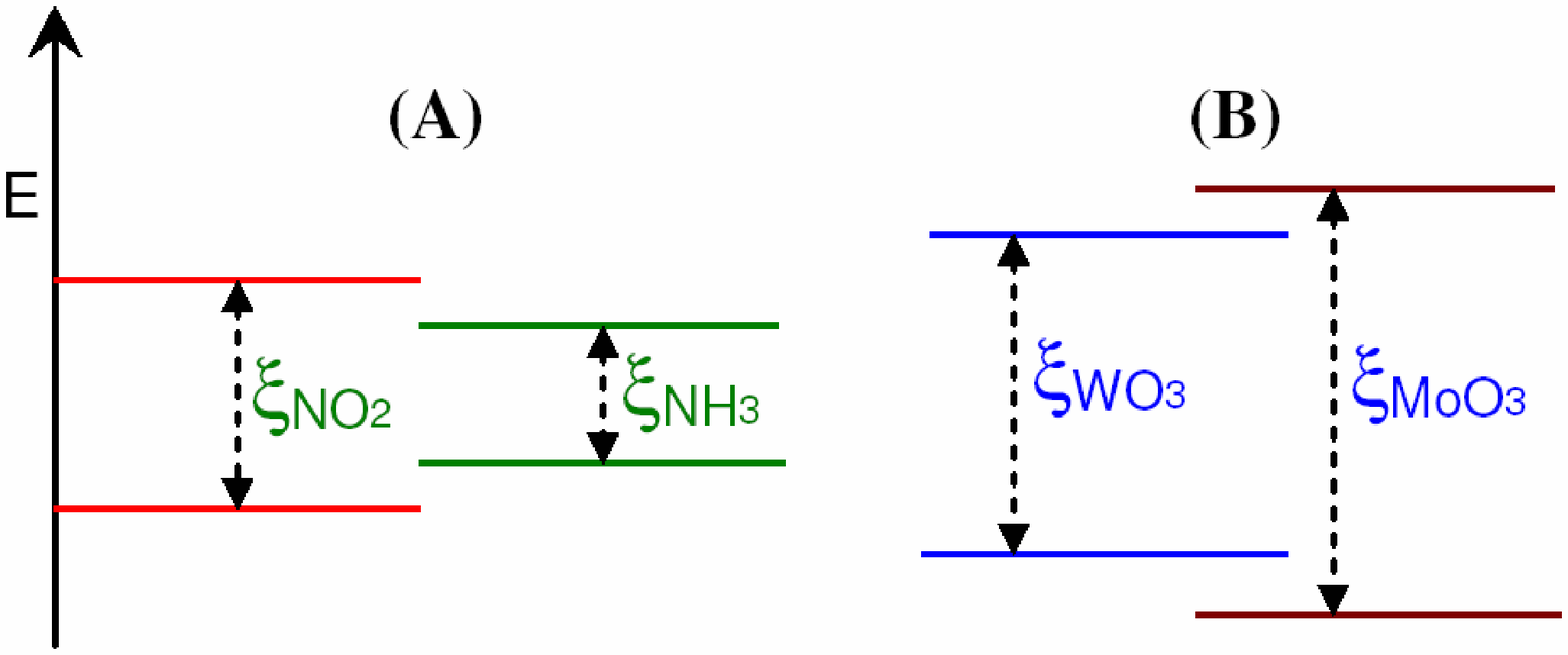}}
\caption{(A) The magnitude of calculated ionization energies for N$^{4+}$O$^{2-}_2$ and N$^{3-}$H$^{+}_3$ gaseous satisfy $\xi_{\rm NO_2} > \xi_{\rm NH_3}$. (B) Whereas, for WO$_3$ and MoO$_3$ solid surfaces, we estimate the ionization-energy inequality as $\xi_{\rm Mo^{6+}O^{2-}_3} (\propto \xi_{\rm Mo^{6+}}) > \xi_{\rm W^{6+}O^{2-}_3} (\propto \xi_{\rm W^{6+}})$ where the proportionalities are due to negligible variation in oxygen concentration. One should not compare the energy levels between the gaseous system (A) and the solid surfaces (B). Again, the energy levels are not to scale and see text for details.}
\label{fig:4}
\end{center}
\end{figure}

Figure~\ref{fig:4}(A) shows the energy-level spacing for gaseous N$^{4+}$O$^{2-}_2$ and N$^{3-}$H$^{+}_3$, while the energy-level spacing in Fig.~\ref{fig:4}(B) are for WO$_3$ and MoO$_3$ solid surfaces. The analysis for these gaseous are somewhat subtle here, but it will be clear once the electron-transfers between ions are understood. For a NO$_2$ molecule, N acts as a cation, while O as an anion, which means that we need to consider the valence state of 4+ since 4 electrons have been transfered from N$^{4+}$ to O$^{2-}_{2}$. This electron-transfer is due to $\alpha_d^{\rm N^{4+}} > \alpha_d^{\rm O^{4+}}$, which is from $\xi_{\rm N^{4+}} < \xi_{\rm O^{4+}}$ (see Table~\ref{Table:I}). In contrast, N is an anion in NH$_3$ molecule due to $\alpha_d^{\rm N^{+}} < \alpha_d^{\rm H^{+}}$ or $\xi_{\rm N^{+}} > \xi_{\rm H^{+}}$ where 3 electrons from 3 hydrogen atoms are transfered to nitrogen, giving N$^{3-}$H$^{+}_3$. Having understood the electron-transfer from the cations to anions based on the ionization energy averaging, we can now show that the polarizability of the molecule NO$_2$ depends on these 4 electrons contributed by nitrogen, and therefore, $\xi_{\rm N^{4+}}$ needs to be compared with $\xi_{\rm H^{+}}$ (3 electrons contributed by 3 hydrogen ions). Obviously, we have $\xi_{\rm N^{4+}} > \xi_{\rm H^{+}}$, which allows us to conclude that $\alpha_d^{\rm NO_2} < \alpha_d^{\rm NH_{3}}$. 

The comparison here is straightforward because (i) we are dealing with molecules that have N as a cation in NO$_2$, and an anion in NH$_3$, and (ii) there are only two types of atoms in each molecule: nitrogen and oxygen in NO$_2$, and nitrogen and hydrogen in NH$_3$. Hence, the ionization energy averaging in multi-element and structurally complicated gas molecules may not be straightforward and one has to be careful in averaging them.


We can now recall the experimental results obtained in Ref.~\cite{pras} for discussion. These results indicate that WO$_3$ is highly sensitive to NO$_2$, even down to 1 part(s)-per-million (ppm) or 0.0001$\%$, while MoO$_3$ is sensitive down to 10 ppm (0.001$\%$) of NH$_3$. Both measurements were carried out at 450$^{\rm o}$C, and these oxides are not doped. Invoking the polarizability-dependent physisorption mechanisms stated earlier, and our conclusion, $\alpha_d^{\rm NO_2} < \alpha_d^{\rm NH_{3}}$, we can surmise here that $\alpha_d^{\rm MoO_3} < \alpha_d^{\rm WO_3}$, which implies that $\xi_{\rm MoO_3} > \xi_{\rm WO_3}$, and therefore $\xi_{\rm Mo^{6+}} > \xi_{\rm W^{6+}}$. Finally, we have concluded that the average ionization energy for W$^{6+}$ is less than Mo$^{6+}$ where there are no experimental values available to calculate the average ionization energy for W$^{6+}$. Our estimation, $\xi_{\rm MoO_3} > \xi_{\rm WO_3}$ is valid given the fact that the band gaps satisfy $E_g^{\rm MoO_3} > E_g^{\rm WO_3}$ as reported in Refs.~\cite{kof,bou}. The band gaps are 2.62 and 3.3 eV for WO$_3$ and MoO$_3$, respectively. As a result, one may exploit the above example to select suitable oxide or other inorganic surfaces as sensors to a given gas molecule, atom or plasma.   

Additional evidences in support of the physisorption mechanisms [given in (a) and (b)] come from the changes to the resistance ($R$) measurements of the oxides, WO$_3$ and MoO$_3$ in different gas environments, NO$_2$ and NH$_3$, respectively. For example, the conductance ($1/R$) is found to increase for MoO$_3$:NH$_3$ system, where larger polarizability of NH$_3$ enhances electron transfer from NH$_3$ to the MoO$_3$ surface, thus increasing the conductance. Therefore, the MoO$_3$:NH$_3$ system satisfies mechanism (a) [see Fig.~\ref{fig:3}(A)]. On the other hand, WO$_3$:NO$_2$ system satisfies mechanism (b), in which, larger polarizability of WO$_3$ oxide surface gives rise to easy electron transfer from WO$_3$ to the NO$_2$ gas molecules [see Fig.~\ref{fig:3}(B)]. This latter electron-transfer will increase the resistance, or decrease the conductance, compared to MoO$_3$:NH$_3$ system. Such resistance measurements were reported in Refs.~\cite{gouma,pras} in agreement with our proposed microscopic mechanisms. This completes our analysis and explanations as to why and how MoO$_3$ and WO$_3$ oxide surfaces are sensitive to NH$_3$ and NO$_2$ gas molecules, respectively.           

Before we conclude, it is also worth to summarize here that the concept of ionization energy approximation was first derived and applied in cuprate superconductors~\cite{and1}. There have been significant theoretical progress since then on our understanding of the IET and its applications where it is still used unambiguously and with high-level self-consistency without \textit{any} patch. This is interesting given the fact that the IET has been successfully used to develop theoretical models to explain the physical properties of different classes of materials. In particular, high temperature superconductors, iron pnictides, ferroelectrics, ferromagnets, diluted magnetic semiconductors, carbon nanotubes, under-screened Kondo metals, alkali halides, multi-element semiconductors and insulators, including nanocrystals and quantum dots~\cite{arul}. 

\section{Conclusions}

We have developed a rigorous and unequivocal strategy to understand the physisorption of NO$_2$ and NH$_3$ molecules on WO$_3$ and MoO$_3$ solid surfaces. Here, we have explained that the highly polarizable molecule can be efficiently detected by a less polarizable solid surface, or vice versa. This knowledge can be readily exploited by experimenters to fine-tune the selection of oxides for sensors to detect certain gaseous, or specifically to detect one type of gas with the highest sensitivity or physisorption. For example, gas molecules with small polarizability can be efficiently detected with oxide surfaces that have large polarizability at a higher operating (surface) temperature.

Current experimental techniques such as direct plasma oxidation (low temperature), thermally grown and followed by annealing (reductive or oxidative ambient) processes and/or plasma deposition techniques~\cite{sankaran,igor,mariotti} can be employed to test our predictions presented here. Furthermore, oxidation states for cations and oxygen ions can be measured accurately with the electron energy loss spectroscopy (EELS). Therefore, the microscopic gas-oxide interactions are now in complete form to be employed by experimenters to design, select and test the new gas-sensing materials for wide applications in the fields of medicine and engineering. 

\section*{Acknowledgments}

This work was supported by the Slovene Human Resources Development and Scholarship Fund (Ad-Futura), the Slovenian Research Agency (ARRS) and the Institut Jo$\check{\rm z}$ef Stefan (IJS). 

%






\end{document}